\documentclass{emulateapj}
\usepackage[varg]{txfonts}
\usepackage{natbib,twoopt}
\bibpunct{(}{)}{;}{a}{}{,}
\pdfoutput=1 

\def\apjl{"Astrophys. J." Lett.}          
\def\apjs{"Astrophys. J. Suppl."}         
\def\aap{"Astron. \& Astrophys."}         

\shorttitle{GCs hosting IMBHs: no mass-segregation based candidates}
\shortauthors{Pasquato et al.}

\begin{document}
\title{Globular clusters hosting intermediate-mass black-holes: no mass-segregation based candidates}

\author{Mario Pasquato}
\affil{Department of Astronomy \& Center for Galaxy Evolution Research, Yonsei University, Seoul 120-749, Republic of Korea}

\author{Paolo Miocchi}
\affil{CNR-Istituto dei Sistemi Complessi, P.le A. Moro 5, 00185 Roma, Italy}

\author{Bong Won Sohn}
\affil{Korea Astronomy and Space Science Institute, 776, Daedeokdae ro, Yuseong gu, Daejeon, 305-348, Republic of Korea}
\affil{Department of Astronomy \& Space Science, University of Science \& Technology, 217 Gajeong-ro, Daejeon, Korea}

\author{Young-Wook Lee}
\affil{Department of Astronomy \& Center for Galaxy Evolution Research, Yonsei University, Seoul 120-749, Republic of Korea}

\begin{abstract}
{Recently, both stellar mass-segregation and binary-fractions were uniformly measured on relatively large samples of Galactic Globular Clusters (GCs). Simulations show that both sizeable binary-star populations and Intermediate-Mass Black Holes (IMBHs) quench mass-segregation in relaxed GCs. Thus mass-segregation in GCs with a reliable binary-fraction measurement is a valuable probe to constrain IMBHs. In this paper we combine mass-segregation and binary-fraction measurements from the literature to build a sample of $33$ GCs (with measured core-binary fractions), and a sample of $43$ GCs (with a binary fraction measurement in the area between the core radius and the half-mass radius). Within both samples we try to identify IMBH-host candidates. These should have relatively low mass-segregation, a low binary fraction ($< 5\%$), and short ($< 1$ Gyr) relaxation time. Considering the core binary fraction sample, no suitable candidates emerge. If the binary fraction between the core and the half-mass radius is considered, two candidates are found, but this is likely due to statistical fluctuations. We also consider a larger sample of $54$ GCs where we obtained an estimate of the core binary fraction using a predictive relation based on metallicity and integrated absolute magnitude. Also in this case no suitable candidates are found. Finally, we consider the GC core- to half-mass radius ratio, that is expected to be larger for GCs containing either an IMBH or binaries. We find that GCs with large core- to half-mass radius ratios are less mass-segregated (and show a larger binary fraction), confirming the theoretical expectation that the energy sources responsible for the large core are also quenching mass-segregation}.
\end{abstract}
\keywords{Methods: statistical - (Galaxy:) globular clusters: general}

\section{Introduction}
Theoretical arguments suggest that Intermediate-Mass Black Holes (IMBHs) may be present in Globular Clusters \citep[GCs;][]{2002MNRAS.330..232C, 2004Natur.428..724P, 2006MNRAS.368..121F}, even though a definitive observational confirmation is still elusive. The presence (or the absence) of IMBHs in GCs would have important implications for cosmology, especially for the formation of Super-Massive black Holes \citep[SMBHs; e.g. see][]{2001ApJ...562L..19E}, and for gravitational wave detection \citep[][]{2002AAS...201.5707B, 2004ApJ...611.1080W, 2004ApJ...613.1143B, 2004ApJ...616..221G, 2008ApJ...681.1431M, 2013A&A...557A.135K}. A promising indirect method for detecting IMBHs in GCs is based on their effect on mass-segregation: in GCs, IMBHs are expected to spend most of their time in a binary with other massive objects, such as stellar-mass black holes, thus injecting energy in the GC core and quenching stellar mass-segregation \citep[][]{2004ApJ...613.1143B, 2007MNRAS.374..857T, 2008ApJ...686..303G}. Primordial binaries behave in a somewhat similar way to an IMBH, also reducing mass segregation dynamically, as shown by \cite{2010ApJ...713..194B}. This leads to an IMBH/binary degeneracy problem in the mass-segregation indicator, which can be solved by measuring the core binary fraction independently. The interplay between mass-segregation and the binary fraction measured in the GC core is further complicated by the fact that mass-segregation may lead to an increased binary fraction in the GC core because binaries are heavier than single stars and thus tend to sink to the center.
 The radial mass-segregation profile was compared to N-body simulations to rule out an IMBH in NGC 2298 by \cite{2009ApJ...699.1511P}, while in M10, instead, mass-segregation data would have been compatible with an IMBH if the core binary fraction were below $\approx 3\%$ \citep[][]{2010ApJ...713..194B}, but this was later shown not to be the case \citep[][]{2011ApJ...743...11D}. 
Recently, \cite{2013ApJ...778...57G} used star counts to derive a uniform measure of mass-segregation by comparing the core radii of \cite{1966AJ.....71...64K} models fit to stars in different mass-bins, over a sample of $54$ GCs. Star counts are not affected by the large fluctuations introduced by the relatively few, luminous stars that dominate surface-brightness measurements, and make it possible to measure mass-segregation in a cluster by comparing the radial distribution of stars of different masses.
Photometric binary fractions for a sample of $59$ GCs from \cite{2012A&A...540A..16M}, based on uniform HST ACS/WFC photometry \citep{2007AJ....133.1658S, 2008AJ....135.2055A}, are also available, resulting in a combined sample of $33$ GCs where both core binary-fractions and mass-segregation are measured, and in a combined sample of $43$ GCs for which mass-segregation and binary-fractions measured between the core and the half-mass radius are available.
In this paper we use this information to identify clusters that:
\begin{itemize}
\item are dynamically old, with a relaxation time $< 1$ Gyr,
\item have low mass-segregation (based on criteria discussed in the following), and
\item have a binary-fraction $< 5\%$.
\end{itemize}
These would be candidates for more in-depth testing, either by a tailored application of the mass-segregation method or by more direct approaches, such as radial velocity and proper motion searches. However, we fail to identify strong candidates. This may be due to shortcomings of our sample or to a genuine lack of GCs where IMBHs are responsible for mass-segregation quenching in the absence of a large binary fraction.

\section{The dataset}
\label{dataset}
 \cite{2013ApJ...778...57G} measured the mass-segregation of main-sequence stars in $54$ Milky Way GCs by fitting \cite{1966AJ.....71...64K} models to star counts binned in stellar mass. They found a simple law in the form
\begin{equation}
r_0 = A \times {\left ( \frac{M}{M_\odot} \right)}^B
\end{equation}
where $r_0$ is the scale radius of the \cite{1966AJ.....71...64K} model fitting stars of mass $M$, and $A$ and $B$ are two parameters. The parameter $A$ is the scale radius of solar-mass stars. The parameter $B$ is a measure of mass-segregation: if it were $0$, all the stars would be distributed equally, independent of mass, while for negative values, heavier stars have a smaller scale radius.
So in order to measure mass-segregation we adopted the $B$ parameter from \cite{2013ApJ...778...57G}, Table 2. 

We {adopt} photometric binary fractions for a sample of $59$ GCs from \cite{2012A&A...540A..16M}, based on uniform HST ACS/WFC photometry \citep{2007AJ....133.1658S, 2008AJ....135.2055A}. In particular we adopted the total binary fraction in the core (last column of their Table 2, $r_C$ sample) and the total binary fraction in the area comprised between the core and the half-mass radius (last column of their Table 2, $r_{C-HM}$ sample).

{We obtained the half-mass relaxation time from \cite{1996AJ....112.1487H}, and the core- to half-mass radius ratio from \cite{2013ApJ...774..151M}. We also considered ratios between the $A$ parameter and the half-mass radius from \cite{1996AJ....112.1487H} when a core- to half-mass radius ratio was not available in \cite{2013ApJ...774..151M}.} We use the $A$ parameter instead of the \cite{1996AJ....112.1487H} core radius because we favour star-count based indicators, as shot noise due to bright stars negatively affects surface-brightness-based indicators. This issue impacts the core radius much more than the half-mass/half-light radius.
{We have assigned a symbol to each quantity we considered in order to keep a consistent and compact notation throughout tables and figures. A quick-look table for the adopted notation is provided in Tab.~\ref{table:0}.}

\begin{table}
\caption{{Summary of the adopted parameters (Col. 1) and our notation (Col. 2). The number of GCs also in the mass-segregation sample for each parameter is reported in Col. 3. References in Col. 4: 1$-$~\cite{2012A&A...540A..16M}, 2$-$~\cite{1996AJ....112.1487H}; 2010 edition, 3$-$~\cite{2013ApJ...774..151M}}}             
\label{table:0}      
\centering                                     
\begin{tabular}{l c c c}         
\hline\hline                        
 Quantity & Symbol & Sample size & Reference \\    
\hline                                   
    Core binary fraction & $f_C$ & $33$ & 1\\ 
    Core-half-mass binary fraction & $f_{C-HM}$ & $43$ & 1\\ 
    Log half-mass relax. time & $\log T_h$ & $54$ & 2\\
    Core- to half-mass-radius ratio & $R_c/R_e$ & $54$  & 3,2\\
\hline                                             
\end{tabular}
\end{table}

Additionally, in order to extend our study to the largest possible number of GCs, i.e. to the whole sample with a measure of mass-segregation by \cite{2013ApJ...778...57G}, we derived an empirical relation to predict $f_C$ as a function of the cluster's integrated absolute magnitude $M_V$ and its metallicity $[Fe/H]$. In this way we can fill in the values of $f_C$ for all the $54$ clusters in \cite{2013ApJ...778...57G}. The relation was obtained by linear regression over the sample of $36$ clusters with a measured $f_C$ from \cite{2012A&A...540A..16M}. We used metallicity and magnitude values from \cite{1996AJ....112.1487H}, which are available for all the clusters in the sample from \cite{2012A&A...540A..16M}. The best fit relation we obtained is
\begin{equation}
\label{relatio}
f_C = 0.55 + 0.05 (M_V + [Fe/H])
\end{equation}
with a standard deviation of residuals (over the dataset used for its derivation) of $0.05$. The scatter is driven mainly by clusters with large $f_C$, while the relation is tighter for the low $f_C$ regime we are interested in (see Fig.~\ref{relatiofig}). This relation was obtained empirically by looking for parameters that correlate with $f_C$ on the \cite{2012A&A...540A..16M} sample, but is likely to reflect regularities of the underlying physics of binary formation and evolution in GCs. \cite{2012A&A...540A..16M} already pointed out that absolute magnitude and binary fraction correlate in their sample, and suggested an explanation based on theoretical models \citep[][]{2008MNRAS.388..307S, 2009ApJ...707.1533F}. \cite{2008MNRAS.388..307S} predicts that binary ionization efficiency is proportional to cluster mass, so that higher magnitude (lower mass) GCs are less efficient in destroying binaries dynamically. On the other hand, that metallicity may influence binary fractions through the cross-section for binary formation via tidal capture was suggested by \cite{1995ApJ...439..687B} and \cite{2006ApJ...636..979I} in the context of low-mass X-ray binary studies.

\section{Results}
\begin{figure}
\includegraphics[width=0.95\columnwidth]{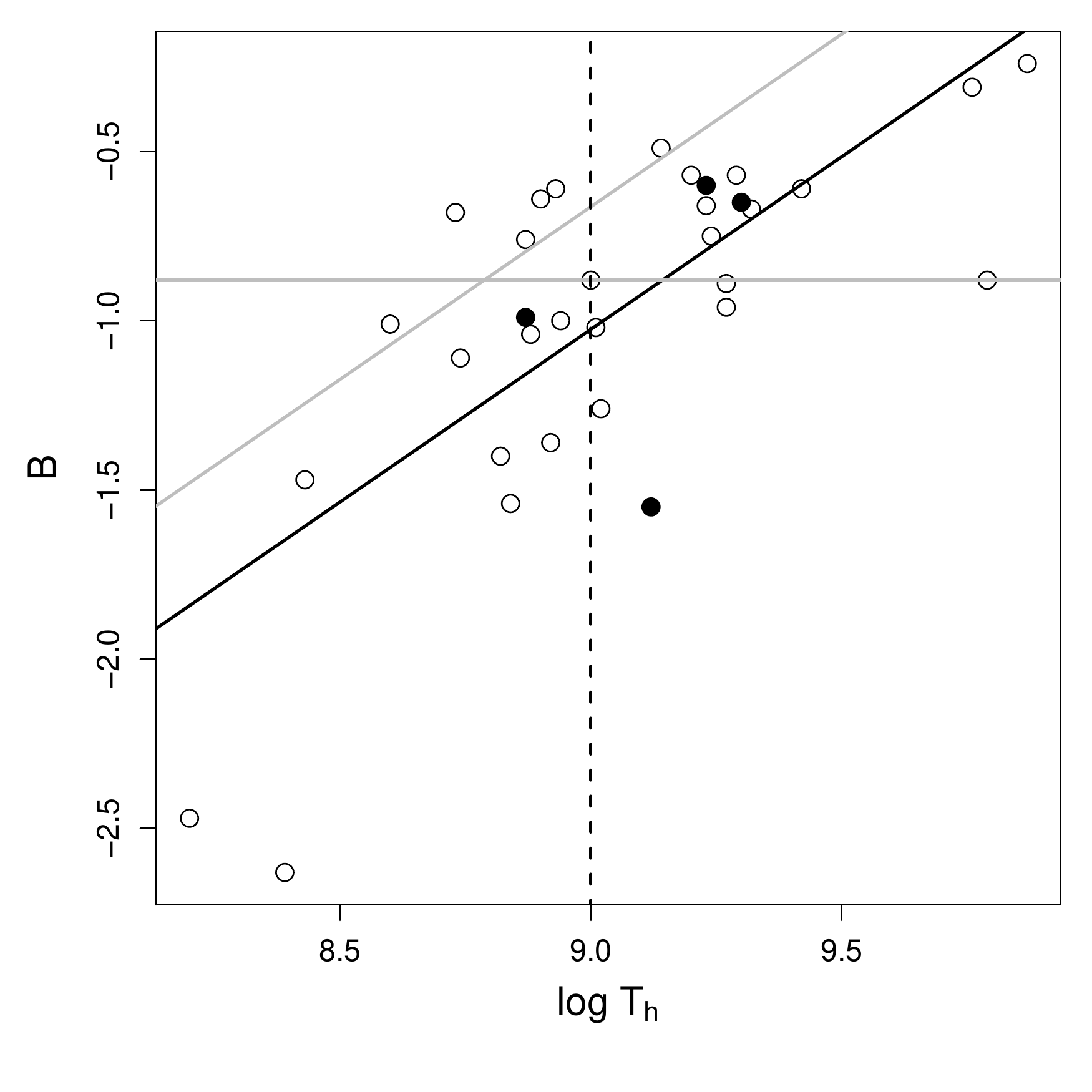}
\caption{{Mass-segregation parameter $B$ as a function of the log half-mass relaxation time. Empty circles are GCs with a total core binary fraction \citep[from][]{2012A&A...540A..16M} over $5\%$, and filled circles below $5\%$. According to the dynamical arguments discussed in \cite{2008ApJ...686..303G} and \cite{2009ApJ...699.1511P}, a good candidate for hosting an IMBH would have low mass-segregation despite being dynamically old, in the absence of a sizeable core binary population. If present in this sample, such a candidate would be represented by a filled circle lying in the upper-left corner in this plot, but there is none. The black solid line is a linear least-square fit, the oblique gray solid line is one-sigma above the best-fit, the horizontal gray solid line is the median of $B$, and the dashed line represents the  boundary (arbitrarily chosen at $1$ Gyr) between relaxed (on the left side) and non-relaxed (on the right side) clusters.}\label{logthrc}}
\end{figure}

\begin{figure}
\includegraphics[width=0.95\columnwidth]{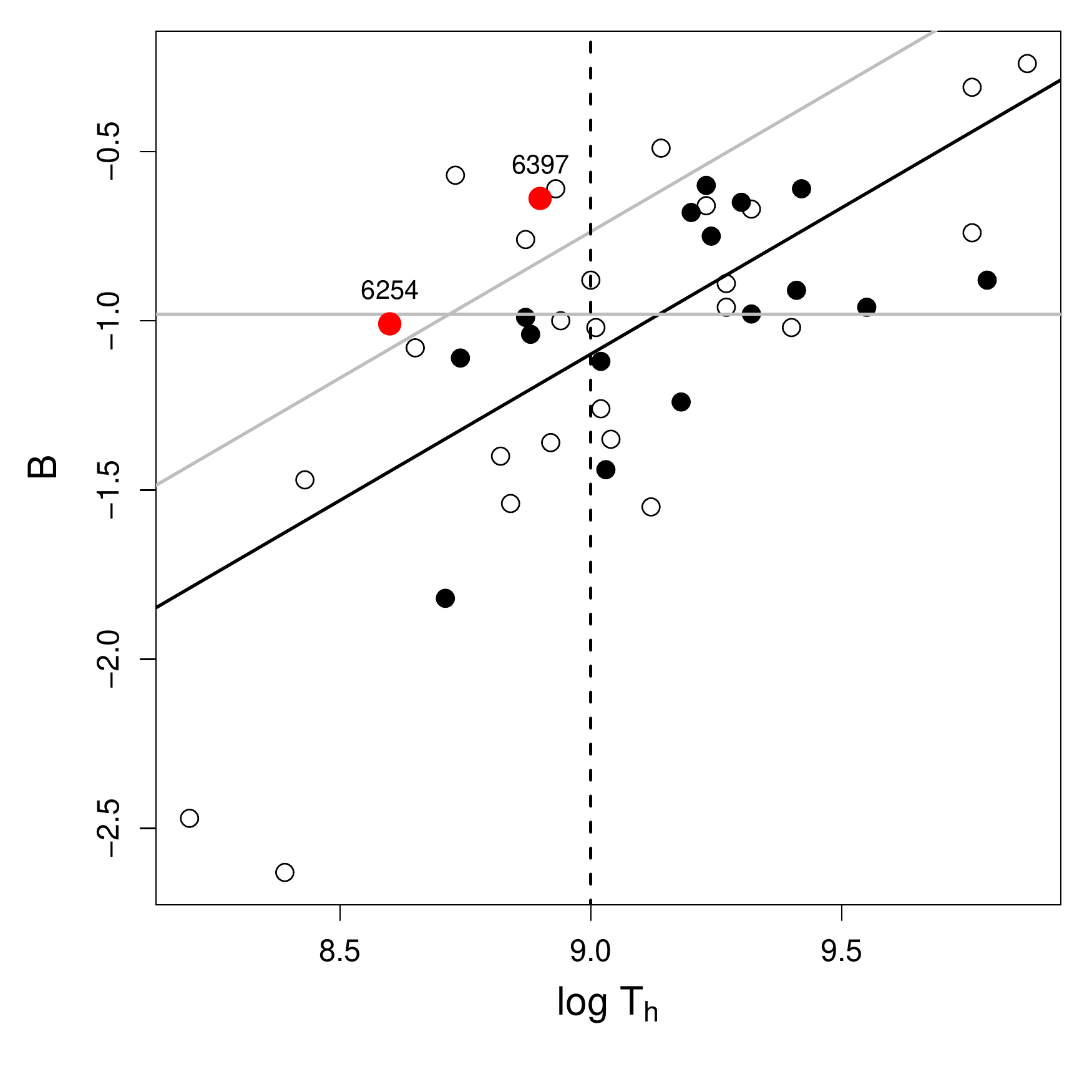}
\caption{{Same as Fig.~\ref{logthrc}, but using the total binary fraction \citep[from][]{2012A&A...540A..16M} in the region comprised between the core and the half-mass radius to label clusters instead of the core binary fraction. Candidates with binary fraction under $5\%$ and at least $1-$sigma away from the best fit line for mass-segregation as a function of relaxation time are represented by a large red filled circle.}\label{logth}}
\end{figure}

In Fig.~\ref{logthrc} we plot the mass-segregation parameter $B$ as a function of the log half-mass relaxation time, dividing the GCs in core binary-fraction ($f_C$) bins. In this plot a strong IMBH-host candidate would be a relaxed GC (i.e. a GC with short relaxation time, less than $1$ Gyr), with a low $f_C$ and low mass-segregation. Such a GC would lie in the upper-left corner of Fig.~\ref{logthrc}, and be represented by a filled circle. The upper-left corner of the figure is however devoid of filled circles, suggesting that in this sample we do not have a clear cut situation where binaries can be excluded and IMBHs are left as the only plausible cause for low observed mass-segregation.
Considering $f_{C-HM}$ instead of $f_C$, we obtain Fig.~\ref{logth}. Quantitatively, a candidate can be defined as being relaxed ($\log T_h < 9$, left of the dashed line), having $f_{C-HM} < 0.05$ (filled circles), and lying one sigma above the best fit regression line for mass-segregation as a function of relaxation time, thus being less segregated than expected based on its relaxation time. Given the relatively large number of GCs with $f_{C-HM} < 0.05$ ($18$, as opposed to $4$ with $f_C < 0.05$), it is unsurprising that two GCs match this criterium. They are represented by filled, slightly bigger, red circles. They are NGC 6397 and NGC 6254. These GCs are also part of the $f_C$ sample, but have in all cases $f_C > 0.05$. Were the threshold set at just two sigma, we would again have no candidates even for $f_{C-HM}$. We conclude that there are no strong candidates for IMBH hosts that can be spotted by mass-segregation quenching alone, at least in this sample. This may be due to the fact that we have few GCs with a low binary fraction in our adopted sample. We do not know whether this is a chance occurrence or a systematic selection effect, but the best we can do in both cases is to increase the number of clusters in our sample.

\begin{figure}
\includegraphics[width=0.95\columnwidth]{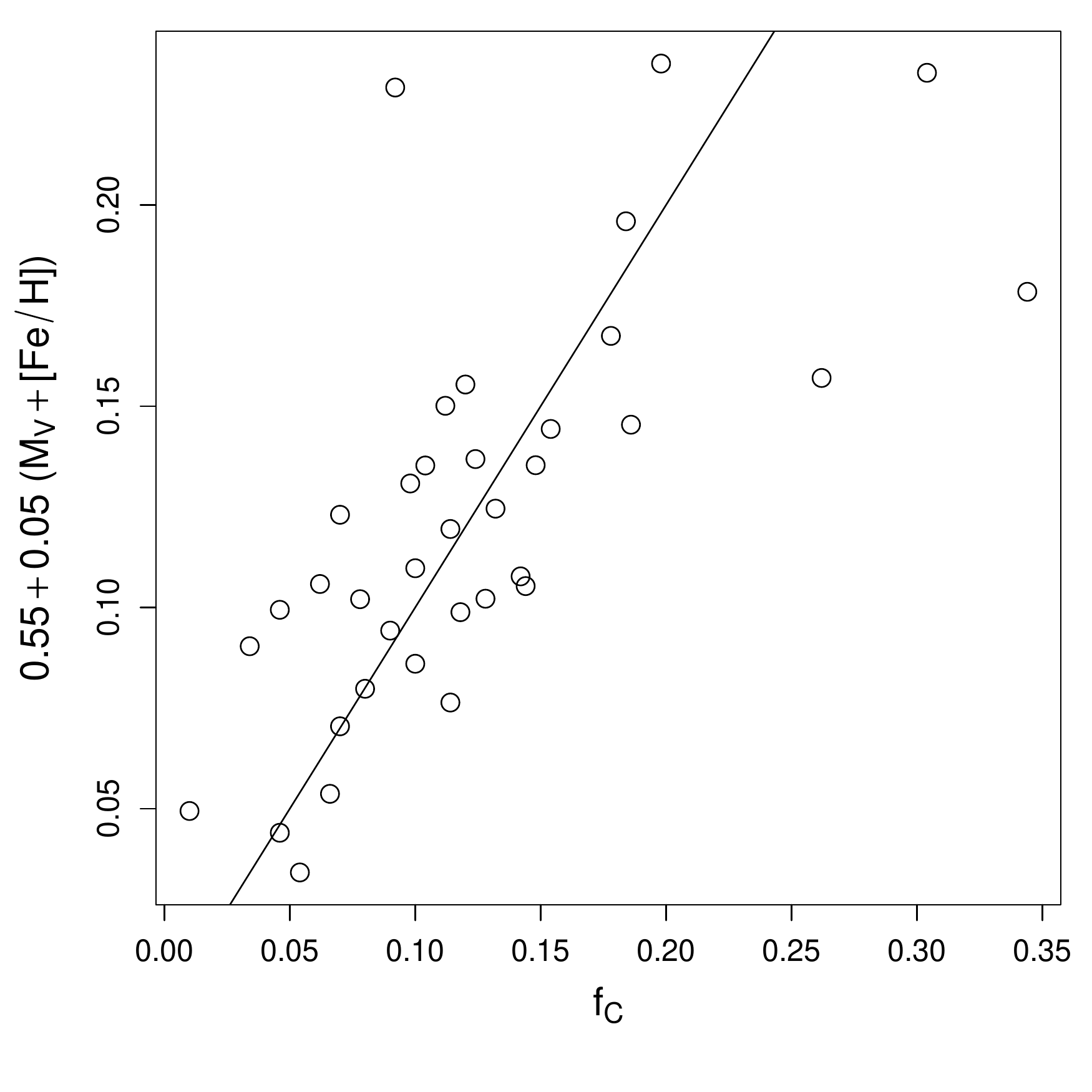}
\caption{Predicted core binary fraction (based on Eq.~\ref{relatio}) as a function of measured $f_C$ on the sample of $36$ GCs used to derive Eq.~\ref{relatio}. The solid line is the identity relationship.\label{relatiofig}}
\end{figure}

\begin{figure}
\includegraphics[width=0.95\columnwidth]{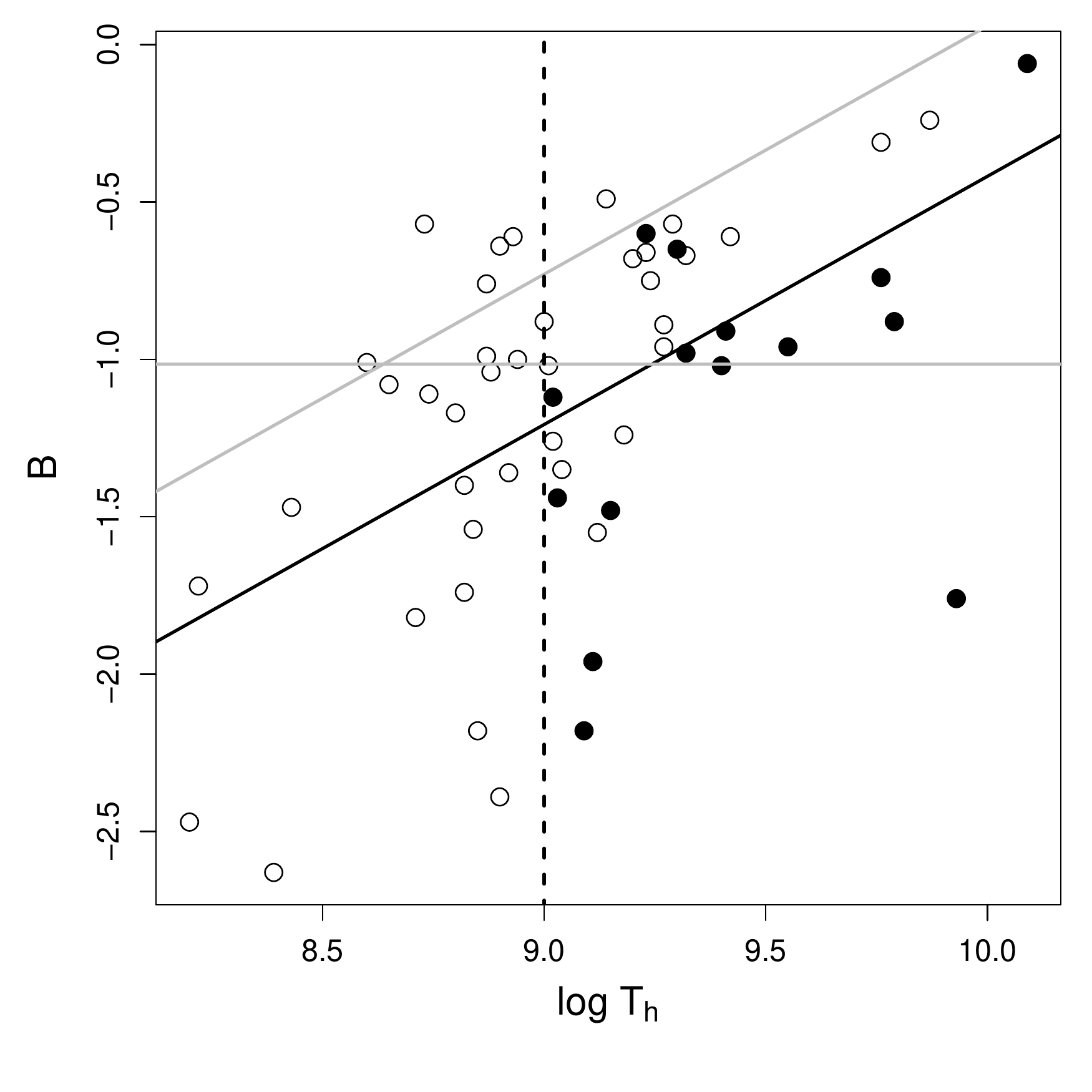}
\caption{{Same as Fig.~\ref{logthrc}, but using the estimated core binary fraction from Eq.~\ref{relatio} on the whole \cite{2013ApJ...778...57G} sample. Also in this case there are no candidates with binary fraction under $5\%$ and at least $1-$sigma away from the best fit line for mass-segregation as a function of relaxation time.}\label{logth_predbinf}}
\end{figure}

Therefore, we considered the full sample of $54$ clusters with a measurement of $B$ from \cite{2013ApJ...778...57G}, by using estimated values of $f_C$ based on Eq.~\ref{relatio}. While the scatter on that relationship is relatively large, as can be seen in Fig.~\ref{relatiofig}, it is still a sufficiently good approximation for the purposes of our paper. We show the results obtained on this larger sample in Fig.~\ref{logth_predbinf}. Also in this case no suitable candidates for hosting an IMBH emerge, i.e. there is no GC with $f_C < 0.05$ that deviates more than 1-sigma from the mass-segregation VS relaxation time relationship in the direction of low mass-segregation. Actually GCs with $f_C < 0.05$ appear to be located systematically below the best fit relation represented by the solid line in Fig.~\ref{logth_predbinf}, suggesting that in the absence of a large binary fraction GCs tend to undergo a larger amount of mass segregation for a given dynamical age. 

\subsection{{Relaxation and energy sources in the core}}
\begin{figure}
\includegraphics[width=0.95\columnwidth]{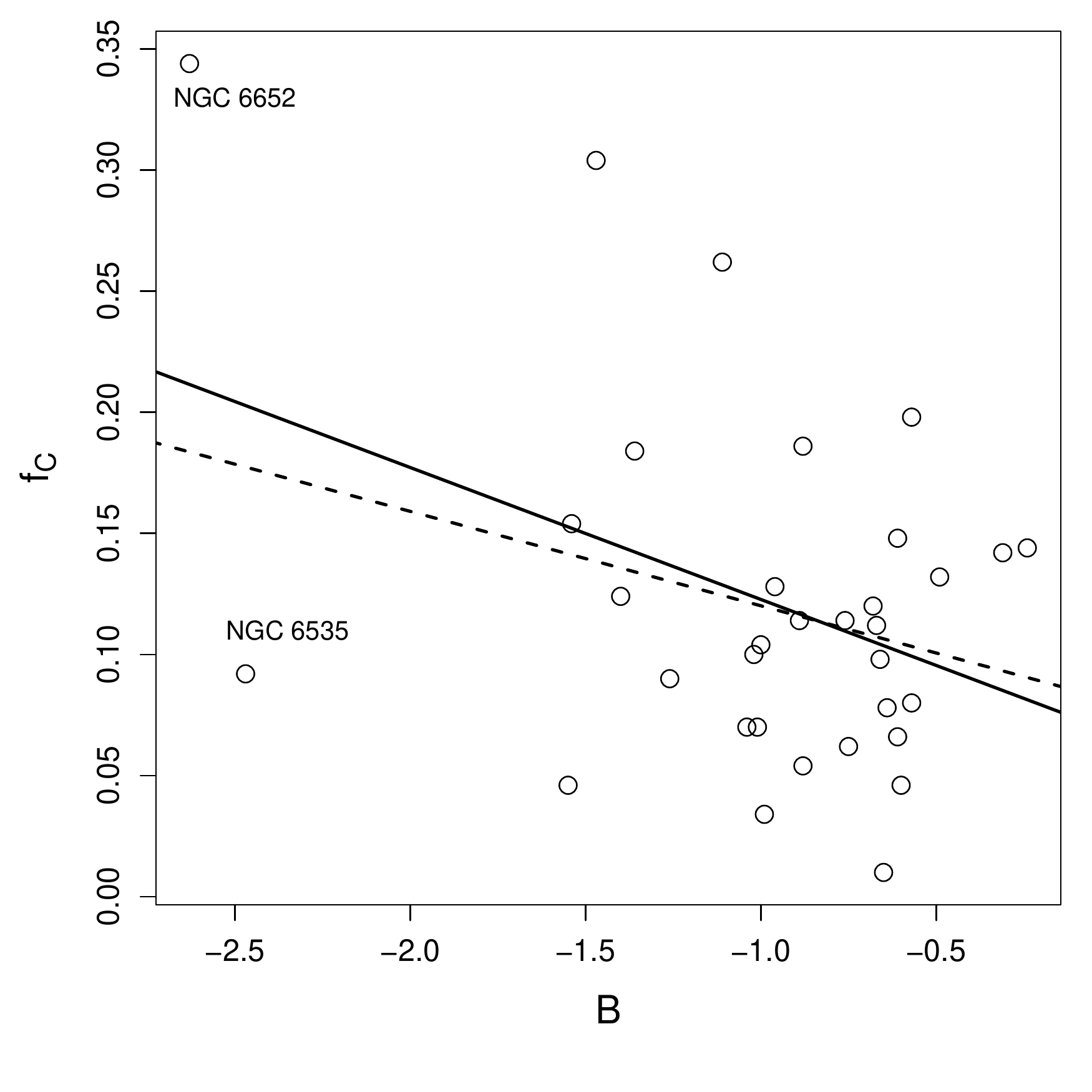}
\caption{{Core binary-fraction measured photometrically by \cite{2012A&A...540A..16M} as a function of the mass-segregation parameter $B$. The solid line is the least-square linear regression. Two outliers are identified and the regression is re-run without them, resulting in the dashed line. Mass-segregated clusters tend to have higher binary-fractions in the core, likely due to mass-segregation of the binaries.} \label{bin}}
\end{figure}

We also find a correlation between the core binary fraction and mass-segregation, i.e. that more mass-segregated clusters have a larger binary fraction in their cores, as shown in Fig.~\ref{bin}. The correlation is expected, because binaries are heavier than single stars and tend to segregate to the core, so that core binary fractions are understandably higher in clusters more affected by mass-segregation.

\begin{figure}
\includegraphics[width=0.95\columnwidth]{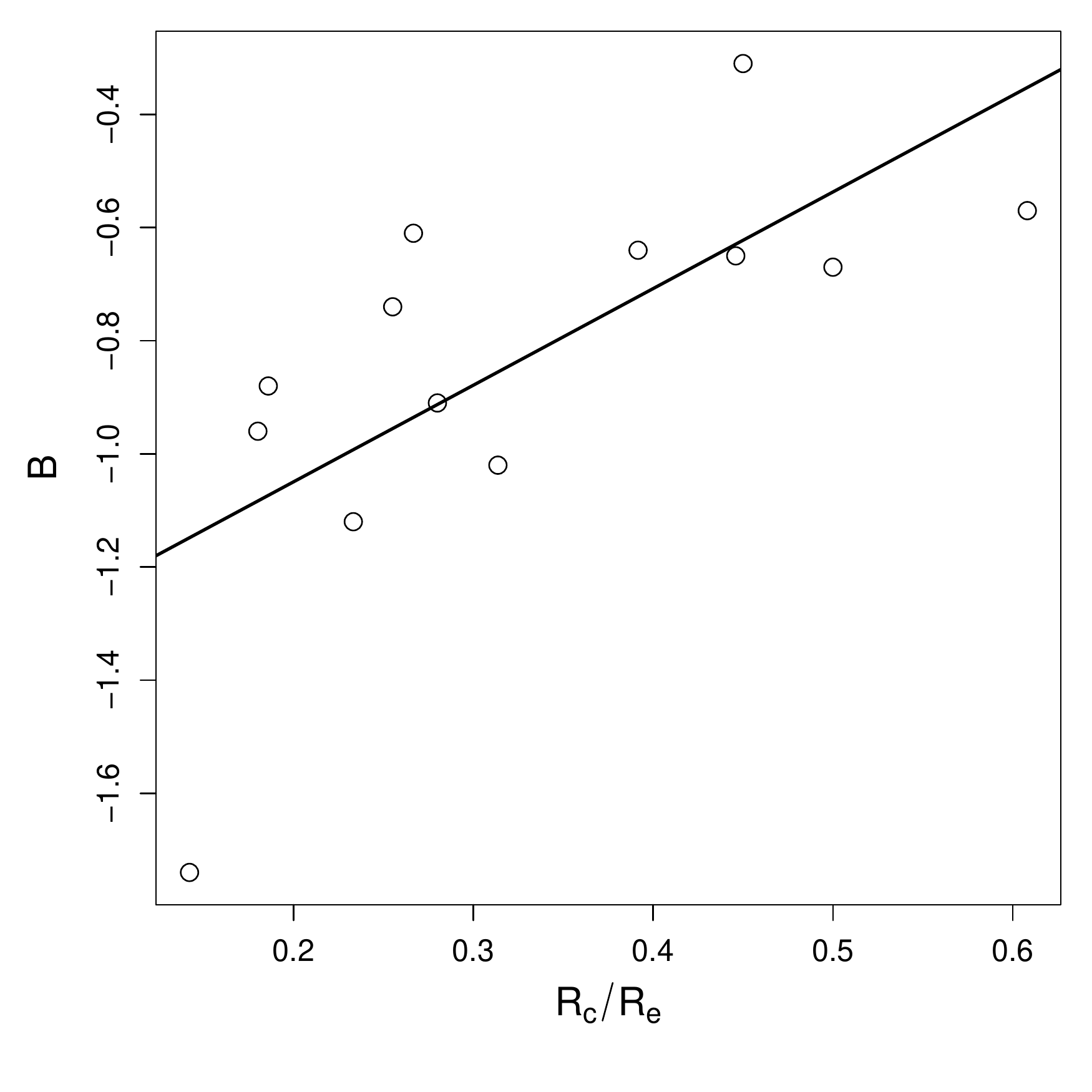}
\caption{Mass-segregation parameter $B$ as a function of the core to half-mass radius ratio from \cite{2013ApJ...774..151M}.\label{mio}}
\end{figure}

\begin{figure}
\includegraphics[width=0.95\columnwidth]{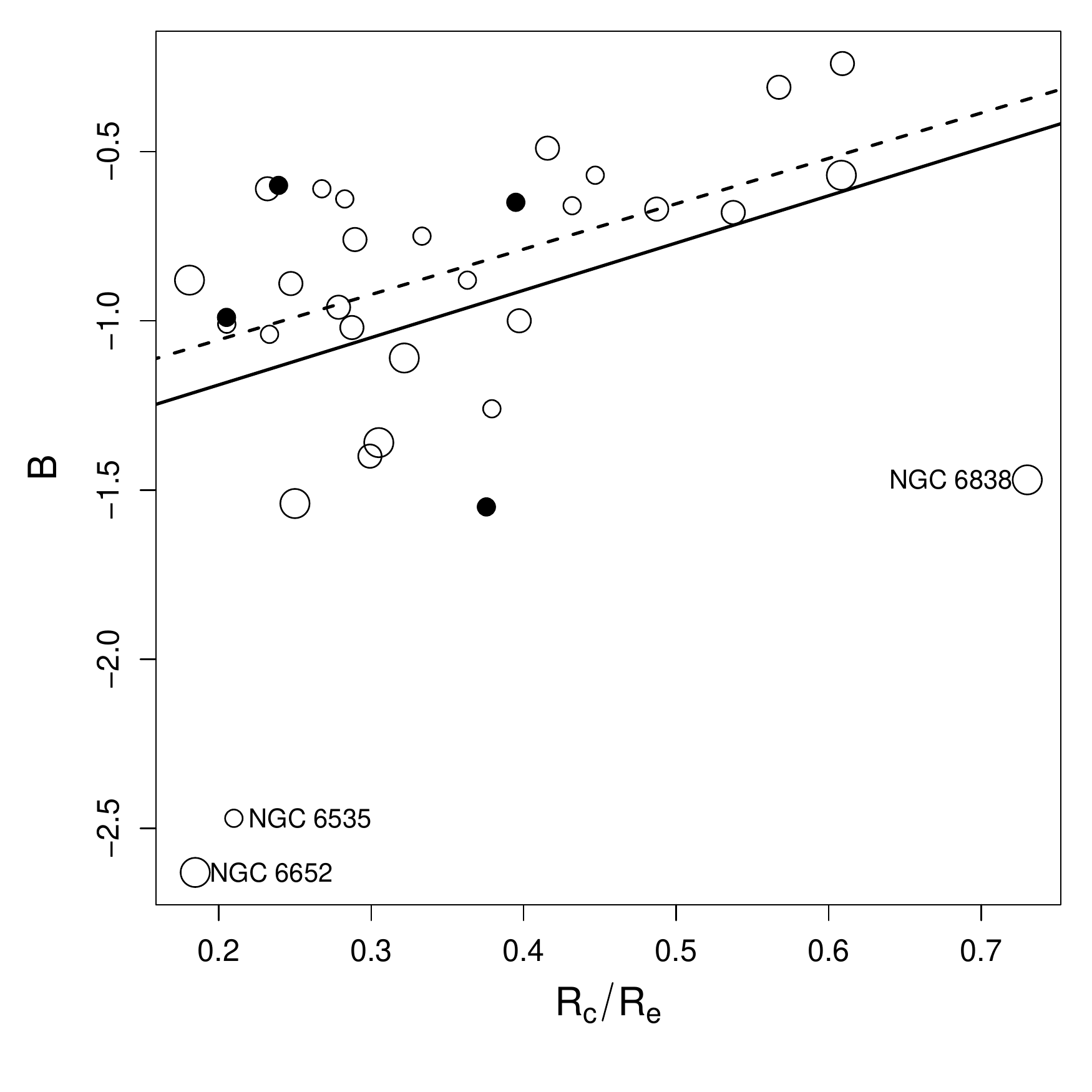}
\caption{Mass-segregation parameter $B$ as a function of the ratio between the \cite{1966AJ.....71...64K} model scale radius of $1$ $M_\odot$ stars, i.e. \cite{2013ApJ...778...57G} $A$ and the half-mass radius ratio from \cite{1996AJ....112.1487H}. The solid line shows the linear fit including all points, while the dashed line excludes the outliers (marked with their NGC number in the plot). Filled circles are GCs with a total core binary fraction \citep[from][]{2012A&A...540A..16M} below $5\%$, empty circles are the remaining GCs. The size of the empty circles increases with their binary fraction. GCs with a lower binary fraction tend to have a smaller core, so they lie towards the left of the plot. Clusters with a large core (with respect to the half-mass radius) and low mass-segregation (i.e. those lying in the upper-right corner of the plot) despite a low binary fraction would be candidates for hosting an IMBH. However no such clusters are found on this plot, as all clusters in the upper right corner of the plot have a core binary fraction exceeding $10\%$.\label{mscorebin}}
\end{figure}

{\cite{2013ApJ...774..151M} show that the ratio of GC core- to half-mass radius correlates with an indicator of dynamical age derived from the BSS radial distribution, which is interpreted in terms of mass-segregation}. {It is therefore no surprise that the scatter plot of the core- to half-mass radius against the mass-segregation indicator $B$ shown in Fig.~\ref{mio} suggests} that clusters with a large core are less mass-segregated, because of their younger dynamical age. Binary stars also are expected to play an important role in determining the dynamical evolution of the core, but, unfortunately, the subsample of clusters with a measured binary fraction by \cite{2012A&A...540A..16M} within the \cite{2013ApJ...774..151M} sample is too small {($n = 13$)} to divide into binary-fraction bins. {So we extended the sample to $n = 33$ (i.e. the clusters for which both $B$ and $f_C$ are available) by calculating the core- to half-mass radius ratio using  the $A$ parameter from \cite{2013ApJ...778...57G} (in place of the core radius) and \cite{1996AJ....112.1487H} half-mass radii, which are available for all clusters. On this sample we show the relation of $A/R_e$ (which we still denote as $R_c/R_e$ in the figure for consistency)} with the mass-segregation parameter in Fig.~\ref{mscorebin}. The relations between mass-segregation and core- to half-mass radius ratio still holds, except for few outliers with extremely high mass segregation. There are, instead, no outliers with low mass segregation, which {may be} candidates for hosting an IMBH, especially in combination with a low binary fraction. Clusters with binary fractions below $5\%$ (filled circles Fig.~\ref{mscorebin}) instead appear to generally fit the overall trend and tend to have small cores. 

\section{Conclusions}
\label{conclu}
In this paper we considered the uniform measure of stellar mass-segregation in GCs obtained by \cite{2013ApJ...778...57G} and the core binary fraction ($f_C$) and the binary fraction measured between the core and the half-mass radius ($f_{C-HM}$) by \cite{2012A&A...540A..16M}. We find that:
\begin{itemize}
\item as expected, mass segegation and relaxation time are anticorrelated, with non-segregated GCs usually having longer relaxation times (i.e. being dynamically young),
\item the few outliers to this trend tend to be more mass-segregated than expected based on their dynamical age,
\item those GCs that are, instead, slightly less mass-segregated than expected based on their dynamical age all have a core binary fraction $f_C > 0.05$, consistent with the binaries being responsible for the reduced mass-segregation, both on a sample of $33$ GCs with measured $f_C$ and on an extended sample of $54$ GCs where we estimated $f_C$ by means of a linear relationship with metallicity and total absolute magnitude,
\item we find two clusters that have $f_{C-HM} < 0.05$ and are over one sigma less mass-segregated with respect to the dynamical age expectation: NGC 6397 and NGC 6254. This finding is compatible with a statistical fluctuation and probably does not indicate that these clusters contain an IMBH,
\item the binary-fraction $f_C$ is correlated with mass-segregation, with GCs that are very segregated having a large binary fraction in the core,
\item mass-segregation anticorrelates with the ratio of core- to half-mass radius measured by \cite{2013ApJ...774..151M}, confirming that the energy sources (binaries, segregation of dark remnants, or, potentially, IMBHs) that bring about the swelling of the core also inhibit mass-segregation, as expected theoretically \citep[see e.g.][]{2007MNRAS.374..857T}.
\end{itemize}
Therefore, we conclude that the samples we considered do not include any GC that qualifies as a strong candidate for hosting an IMBH, based on their core binary-fraction. The reason for this is that core binary fractions $f_C$ are high enough, in relaxed clusters that display low mass-segregation, to be responsible for the low mass-segregation observed. This may be due to low-$f_C$ clusters being underrepresented in our adopted sample due to selection effects, but is also confirmed on the larger sample of all clusters with a measured mass-segregation from \cite{2013ApJ...778...57G} when we use an estimated $f_C$. The consistency of our result over this extended sample casts doubts over selection effects playing a significant role in our negative finding.

\section*{Acknowledgements}
This paper is a result of the collaborative project between Korea Astronomy and Space Science Institute and Yonsei University through DRC program of Korea Research Council of Fundamental Science and Technology (DRC-12-2-KASI). M.P. acknowledges support from Mid-career Researcher Program (No. 2015-008049) through the National Research Foundation (NRF) of Korea. P. M. was supported by the Cosmic-Lab project (http://www.cosmic-lab.eu) funded by the European Research Council (contract ERC-2010-AdG-267675). Y.-W. L. acknowledges support from the National Research Foundation of Korea to the Center for Galaxy Evolution Research (No. 2010-0027910). We wish to thank the referee for his/her constructive criticism and comments.
M. P. wishes to thank Prof. Jay Strader, Prof. Laura Chomiuk, and Dr. Evangelia Tremou of Michigan State University for reading and commenting on an early draft of this paper.

\bibliography{manuscript}

\begin{thebibliography}{26}
\expandafter\ifx\csname natexlab\endcsname\relax\def\natexlab#1{#1}\fi

\bibitem[{{Anderson} {et~al.}(2008){Anderson}, {Sarajedini}, {Bedin}, {King},
  {Piotto}, {Reid}, {Siegel}, {Majewski}, {Paust}, {Aparicio}, {Milone},
  {Chaboyer}, \& {Rosenberg}}]{2008AJ....135.2055A}
{Anderson}, J., {Sarajedini}, A., {Bedin}, L.~R., {et~al.} 2008, \aj, 135, 2055

\bibitem[{{Baumgardt} {et~al.}(2004){Baumgardt}, {Makino}, \&
  {Ebisuzaki}}]{2004ApJ...613.1143B}
{Baumgardt}, H., {Makino}, J., \& {Ebisuzaki}, T. 2004, \apj, 613, 1143

\bibitem[{{Beccari} {et~al.}(2010){Beccari}, {Pasquato}, {De Marchi},
  {Dalessandro}, {Trenti}, \& {Gill}}]{2010ApJ...713..194B}
{Beccari}, G., {Pasquato}, M., {De Marchi}, G., {et~al.} 2010, \apj, 713, 194

\bibitem[{{Bellazzini} {et~al.}(1995){Bellazzini}, {Pasquali}, {Federici},
  {Ferraro}, \& {Pecci}}]{1995ApJ...439..687B}
{Bellazzini}, M., {Pasquali}, A., {Federici}, L., {Ferraro}, F.~R., \& {Pecci},
  F.~F. 1995, \apj, 439, 687

\bibitem[{{Bender} \& {Stebbins}(2002)}]{2002AAS...201.5707B}
{Bender}, P.~L. \& {Stebbins}, R.~T. 2002, in Bulletin of the American
  Astronomical Society, Vol.~34, American Astronomical Society Meeting
  Abstracts, 1207

\bibitem[{{Dalessandro} {et~al.}(2011){Dalessandro}, {Lanzoni}, {Beccari},
  {Sollima}, {Ferraro}, \& {Pasquato}}]{2011ApJ...743...11D}
{Dalessandro}, E., {Lanzoni}, B., {Beccari}, G., {et~al.} 2011, \apj, 743, 11

\bibitem[{{Ebisuzaki} {et~al.}(2001){Ebisuzaki}, {Makino}, {Tsuru}, {Funato},
  {Portegies Zwart}, {Hut}, {McMillan}, {Matsushita}, {Matsumoto}, \&
  {Kawabe}}]{2001ApJ...562L..19E}
{Ebisuzaki}, T., {Makino}, J., {Tsuru}, T.~G., {et~al.} 2001, \apjl, 562, L19

\bibitem[{{Fregeau} {et~al.}(2009){Fregeau}, {Ivanova}, \&
  {Rasio}}]{2009ApJ...707.1533F}
{Fregeau}, J.~M., {Ivanova}, N., \& {Rasio}, F.~A. 2009, \apj, 707, 1533

\bibitem[{{Freitag} {et~al.}(2006){Freitag}, {Rasio}, \&
  {Baumgardt}}]{2006MNRAS.368..121F}
{Freitag}, M., {Rasio}, F.~A., \& {Baumgardt}, H. 2006, \mnras, 368, 121

\bibitem[{{Gill} {et~al.}(2008){Gill}, {Trenti}, {Miller}, {van der Marel},
  {Hamilton}, \& {Stiavelli}}]{2008ApJ...686..303G}
{Gill}, M., {Trenti}, M., {Miller}, M.~C., {et~al.} 2008, \apj, 686, 303

\bibitem[{{Goldsbury} {et~al.}(2013){Goldsbury}, {Heyl}, \&
  {Richer}}]{2013ApJ...778...57G}
{Goldsbury}, R., {Heyl}, J., \& {Richer}, H. 2013, \apj, 778, 57

\bibitem[{{G{\"u}ltekin} {et~al.}(2004){G{\"u}ltekin}, {Miller}, \&
  {Hamilton}}]{2004ApJ...616..221G}
{G{\"u}ltekin}, K., {Miller}, M.~C., \& {Hamilton}, D.~P. 2004, \apj, 616, 221

\bibitem[{{Harris}(1996)}]{1996AJ....112.1487H}
{Harris}, W.~E. 1996, \aj, 112, 1487

\bibitem[{{Ivanova}(2006)}]{2006ApJ...636..979I}
{Ivanova}, N. 2006, \apj, 636, 979

\bibitem[{{King}(1966)}]{1966AJ.....71...64K}
{King}, I.~R. 1966, \aj, 71, 64

\bibitem[{{Konstantinidis} {et~al.}(2013){Konstantinidis}, {Amaro-Seoane}, \&
  {Kokkotas}}]{2013A&A...557A.135K}
{Konstantinidis}, S., {Amaro-Seoane}, P., \& {Kokkotas}, K.~D. 2013, \aap, 557,
  A135

\bibitem[{{Mandel} {et~al.}(2008){Mandel}, {Brown}, {Gair}, \&
  {Miller}}]{2008ApJ...681.1431M}
{Mandel}, I., {Brown}, D.~A., {Gair}, J.~R., \& {Miller}, M.~C. 2008, \apj,
  681, 1431

\bibitem[{{Miller} \& {Hamilton}(2002)}]{2002MNRAS.330..232C}
{Miller}, M.~C. \& {Hamilton}, D.~P. 2002, \mnras, 330, 232

\bibitem[{{Milone} {et~al.}(2012){Milone}, {Piotto}, {Bedin}, {Aparicio},
  {Anderson}, {Sarajedini}, {Marino}, {Moretti}, {Davies}, {Chaboyer},
  {Dotter}, {Hempel}, {Mar{\'{\i}}n-Franch}, {Majewski}, {Paust}, {Reid},
  {Rosenberg}, \& {Siegel}}]{2012A&A...540A..16M}
{Milone}, A.~P., {Piotto}, G., {Bedin}, L.~R., {et~al.} 2012, \aap, 540, A16

\bibitem[{{Miocchi} {et~al.}(2013){Miocchi}, {Lanzoni}, {Ferraro},
  {Dalessandro}, {Vesperini}, {Pasquato}, {Beccari}, {Pallanca}, \&
  {Sanna}}]{2013ApJ...774..151M}
{Miocchi}, P., {Lanzoni}, B., {Ferraro}, F.~R., {et~al.} 2013, \apj, 774, 151

\bibitem[{{Pasquato} {et~al.}(2009){Pasquato}, {Trenti}, {De Marchi}, {Gill},
  {Hamilton}, {Miller}, {Stiavelli}, \& {van der Marel}}]{2009ApJ...699.1511P}
{Pasquato}, M., {Trenti}, M., {De Marchi}, G., {et~al.} 2009, \apj, 699, 1511

\bibitem[{{Portegies Zwart} {et~al.}(2004){Portegies Zwart}, {Baumgardt},
  {Hut}, {Makino}, \& {McMillan}}]{2004Natur.428..724P}
{Portegies Zwart}, S.~F., {Baumgardt}, H., {Hut}, P., {Makino}, J., \&
  {McMillan}, S.~L.~W. 2004, \nat, 428, 724

\bibitem[{{Sarajedini} {et~al.}(2007){Sarajedini}, {Bedin}, {Chaboyer},
  {Dotter}, {Siegel}, {Anderson}, {Aparicio}, {King}, {Majewski},
  {Mar{\'{\i}}n-Franch}, {Piotto}, {Reid}, \&
  {Rosenberg}}]{2007AJ....133.1658S}
{Sarajedini}, A., {Bedin}, L.~R., {Chaboyer}, B., {et~al.} 2007, \aj, 133, 1658

\bibitem[{{Sollima}(2008)}]{2008MNRAS.388..307S}
{Sollima}, A. 2008, \mnras, 388, 307

\bibitem[{{Trenti} {et~al.}(2007){Trenti}, {Ardi}, {Mineshige}, \&
  {Hut}}]{2007MNRAS.374..857T}
{Trenti}, M., {Ardi}, E., {Mineshige}, S., \& {Hut}, P. 2007, \mnras, 374, 857

\bibitem[{{Will}(2004)}]{2004ApJ...611.1080W}
{Will}, C.~M. 2004, \apj, 611, 1080

\end{thebibliography}
\bibliographystyle{aa}

\end{document}